\documentclass[aps,amsmath,showpacs,showkeys,superscriptaddress]{revtex4}
\usepackage{color}
\usepackage{amssymb}
\usepackage{graphicx}
\usepackage{bm}
\usepackage{inputenc}
\usepackage{subfigure}
\usepackage{amsmath}
\usepackage{titlesec}
\usepackage{amssymb,amsfonts}
\begin{document}
\title{Integrable systems: From the inverse spectral transform to zero curvature condition}
\author{Basir Ahamed Khan}
\email{basir.khan@gmail.com}
\affiliation{Department of Physics, Krishnath College, Berhampore, Murshidabad 742101, India}
\author{Supriya Chatterjee}
\affiliation{Department of Physics, Bidhannagar College, EB-2, Sector-1, Salt Lake, Kolkata 700064, India}
\author{Sekh Golam Ali} 
\affiliation{Department of Physics, Kazi Nazrul University, Asansol 713303, India}
\author{Benoy Talukdar}
\affiliation{Department of Physics, Visva-Bharati University, Santiniketan 731235, India}
\begin{abstract}
This \textquoteleft research-survey' is meant  for beginners in the studies of integrable systems. Here we outline some analytical methods for dealing with  a class of nonlinear partial differential equations. We pay special attention to   \textquoteleft inverse spectral transform', \textquoteleft Lax pair representation', and \textquoteleft zero-curvature condition' as applied to these equations. We provide a number of interesting exmples to gain some physico-mathematical feeling for the methods presented.
\end{abstract}
\maketitle
\section*{1. Introduction}
Integrable systems are represented by nonlinear partial differential equations (NLPDEs) which, in principle, can be solved by analytic methods. This necessarily implies that the solution of such equations can be constructed using a finite number of algebraic operations and integrations. The inverse scattering method as discovered by Gardner, Greene, Kruskal and Miura \cite{1} represents a very useful tool to analytically solve a class of nonlinear differential equations which support soliton solutions. Solitons are localized waves that propagate without change in their properties (shape, velocity etc.). These waves are stable against mutual collision and retain their identities except for some trivial phase change. Mechanistically, the linear and nonlinear terms in NLPDEs have opposite effects on the wave propagation. In particular, the linear term causes dispersion of the wave while the nonlinear one leads to steepening. In certain equations representing typical physical systems a balance between these two effects produces solitons. Soliton was observed in 1834 by a Scottish engineer, John Scott Russel \cite{2}. He followed the motion of a boat drawn rapidly along a cannel by a pair of horses. The boat suddenly stopped. He observed that a large solitary elevation (well defined heap of water) left the boat and moved with great velocity. The first soliton equation was identified by Zabusky and Kruskal \cite{3} in 1965. By solving the KdV equation numerically, they demonstrated an unexpected property of its solution. It was found that from a smooth initial wave form there emerged waves with sharp peaks which are, in fact, representative of the unexpected phenomena observed by Scott Russel. Solitons appear in a wide variety of physical systems ranging from shallow water waves to Bose-Einstein condensates \cite{4}. Consequently, soliton theory now occupies a large part of theoretical and mathematical physics.
\par In this paper we collect a number of useful results that are needed to construct localized solutions of integrable systems and make appropriate comments wherever necessary. We also pay equal attention to introduce the necessary and sufficient conditions for a general nonlinear evolution equation to be integrable. In presenting the work we have followed the historical development of the subject as far as practicable. It is our belief that for young researchers in this field of investigation, the results presented will serve as an appetizer if not a full meal.
\par  In section 2 we introduce the inverse spectral method for solving the nonlinear evolution equation with particular emphasis on the KdV equation. We use the technique to provide results for both single and multi-soliton solutions. Multi-soliton solutions indicate that at the time $t=0$ we have a single hump produced by juxtaposition of the component solitons and separate components make their appearance for only $t>0$. The taller solitons move faster and carries greater energy compared to that carried by the shorter ones. We provide in section 3 the next important development in the theory of integral equations as developed by Lax \cite{5} who introduced a general principle for associating nonlinear evolution equations with linear differential operators. These linear operators are often called Lax pairs. The Lax pairs yield evolution equations when they commute and represent a useful tool in finding conserved quantities of integrable dynamical systems. We present a number of examples in respect of Lax's treatment of nonlinear partial differential equations. We devote section 4 to realize nonlinear evolution equations as a compatibility condition between pair of linear equations. The idea is to avoid the need to consider higher-order Lax pair and represent nonlinear evolution equation as the zero curvature condition \cite{6}. Here we first briefly discuss as to why the noted compatibility condition has been given the name zero curvature condition and then provide a set of examples to substantiate the ideas of ref. 6. Finally, in section 5 we summarize our outlook on the present work and make some concluding remarks.
\section*{2. Inverse spectral transform for solving the Korteweg-de Vries (KdV) equation}
Traditionally, a class of nonlinear partial differential equations (NLPDEs) is classified as integrable if these can be solved by the use of inverse spectral transform (IST).  The idea of the IST is the following.
\par  Each integrable NPDE is associated with a linear ordinary differential equation (LODE) containing a parameter $\lambda$, usually known as spectral parameter, and the solution $u(x,t)$ of the NPDE appears as a coefficient in the corresponding LODE. The function $u(x,t)$ is known as the potential characterizing the linear problem. In the NLPDE the quantities $x$ and $t$ appear as independent variables. In fact, these are the so-called spatial and temporal coordinates. Here we are concerned with a one-dimensional problem characterized by a single spatial variable. We point out while $x$ is a dynamical variable of the NLPDE, the quantities $\lambda$ and $t$ appear as parameters. One of the most significant properties of the IST is that the spectral parameter does not change with time. The solution of the NLPDE is constructed by using the spectra of the LODE in the Gel'fand-Levitan equation \cite{7}. We shall illustrate the use of IST in solving NPDE with special attention to the well known Korteweg-de Vries  (KdV) equation \cite{8}
\begin{equation}
 u_t-6uu_x+u_{xxx}=0
\end{equation}
where the subscripts of $u$ denote partial derivatives.
\par The Gel'fand-Levitan equation associated with the inverse spectral method is a Volterra integral equation for a function $K(x,y,t)$ written as \cite{1}
\begin{equation}
K(x,y,t)+B(x+y,t)+\int_x^\infty K(x,s,t)B(x+y,t)ds=0 
\end{equation}
with the kernel $B(z,t)$ given by
\begin{equation}
 B(z,t)=\sum_{n=1}^Nc_n^2e^{-\kappa_nz}+\frac{1}{2\pi}\int_{-\infty}^\infty b(k,t)e^{ikz}dk
\end{equation}
In equation (3) $\kappa_n$ and $k$  represent the wave numbers for  bound- and continuum-state energies of the one-dimensional  Schr\"{o}dinger equation with potential $u(x,t)$. Understandably, here $n$ and $b(k,t)$ stand for the principal quantum number and reflection coefficient; $c_n$ is the normalization constant of the bound-state wave function. The solution $K(.)$ of this integral equation is related to the potential by \cite{1}
\begin{equation}
u(x,t)=-2\frac{\partial}{\partial x}K(x,y,t).
\end{equation}
Equation (4) shows that we have obtained the solution of a nonlinear evolution equation by using the solution of a linear ordinary differential equation. We shall now apply this result to obtain the solution of the KdV equation as given in Eq. (1). The spectral problem for the KdV equation is provided by a one-dimensional Schr\"{o}dinger equation given by
\begin{equation}
 H\psi=\lambda\psi.
\end{equation}
The Hamiltonian $H$, in addition to the kinetic energy term, involves $u(x,t)$, the solution of Eq. (1) as the potential energy. We consider the particular case where the potential has a single bound state with the eigenvalue $\lambda=-\kappa^2$, normalization constant $c$, and vanishing reflection coefficient i.e. $b(k,t)=0$. In this case $B(z,t)$  is given by
\begin{equation}
 B(z,t)=c^2(t)e^{-\kappa z}=c_0^2e^{8\kappa^3t}e^{-\kappa z}.
\end{equation}
For $z=x+y$ the kernel of the Gel'fand-Levitan equation is separable such that we can write 
\begin{equation}
 K(x,y,t)+c_0^2e^{8\kappa^3t}e^{-\kappa(x+y)}+c_0^2e^{8\kappa^3t}e^{-\kappa y}\int_x^\infty K(x,s,t)e^{-\kappa s}ds=0.
\end{equation}
Because of separability of the kernel, Eq. (7) can easily be solved to get \cite{9}
\begin{equation}
 K(x,y,t)=-\kappa\frac{e^{8\kappa^3t}e^{-\kappa(y-x_0)}}{\cosh[\kappa(x-x_0-4\kappa^3t)]}
\end{equation}
which when substituted in Eq. (4) gives the solution of the KdV equation in the form 
\begin{equation}
 u(x,t)=-2\kappa^2\sec h^2[\kappa(x-x_0-4\kappa^3t)].
\end{equation}
This solution corresponds to a single soliton of amplitude $-2\kappa^2$ and speed $4\kappa^3$ moving to the right and centered initially at 
\begin{equation}
 x_0=\frac{1}{2\kappa}\ln(\frac{c_0^2}{2\kappa}).
\end{equation}
The method of inverse scattering as formulated above for the KdV equations is an ingenuous way of solving the initial value problem by associating the solution of a nonlinear equation with the potential of a time-independent Schr\"{o}dinger equation. This replaces the problem of finding a solution of the KdV equation - a nonlinear equation- with the less difficult task of solving a linear quantum mechanical scattering problem.
\par In order to gain some physical feeling for the time evolution of the KdV soliton, we plot in Figs. 1- 4  the solution in Eq. (9) at different values of $t$ as a function of $x$ for $x_0=0$. As expected from the formula in Eq. (9), the solitons in Figs. 1 - 4 have centres  at $x=0$, 2, 4 and 6 respectively. In these figures coordinates of the peaks of the solitons are given by (0,2), (2,2), (4,2) and (6,2).
\begin{figure}[!tbp]
 \centering
\begin{minipage}[b]{0.4\textwidth}
\includegraphics[width=\textwidth]{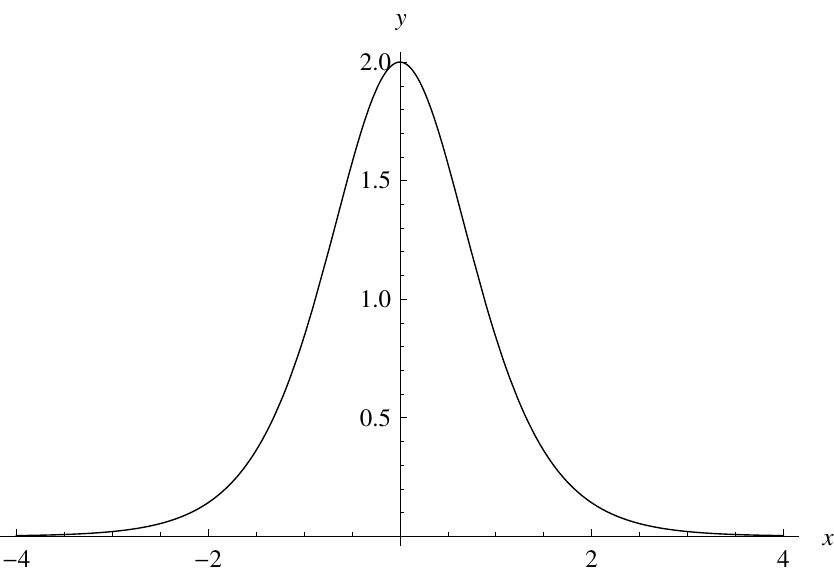}
 \caption{Soliton solution for $\kappa=1$, $t=0$, $x_0=0$, $y=-u(x,0)$}
\end{minipage}
\hfill
\begin{minipage}[b]{0.4\textwidth}
\includegraphics[width=\textwidth]{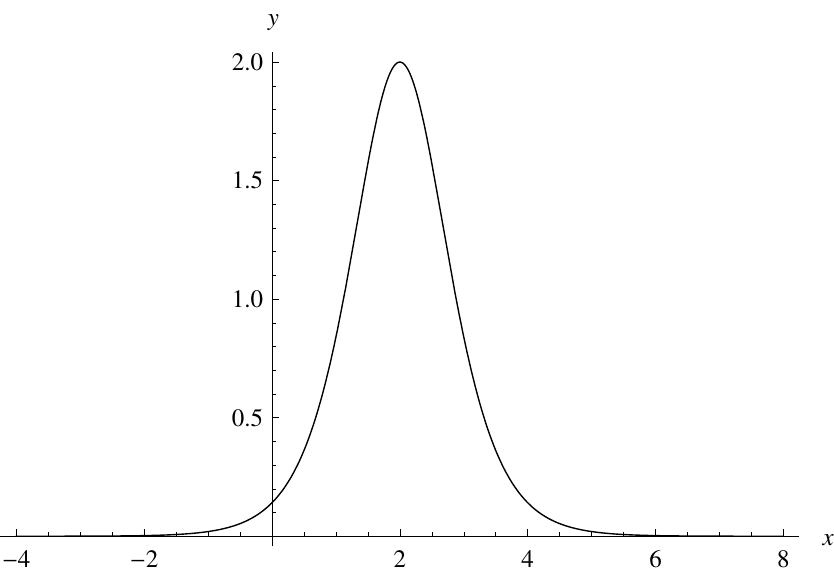}
 \caption{Soliton solution for $\kappa=1$, $t=0.5$, $x_0=0$, $y=-u(x,0.5)$}
\end{minipage}
\end{figure}
\begin{figure}[!tbp]
 \centering
\begin{minipage}[b]{0.4\textwidth}
\includegraphics[width=\textwidth]{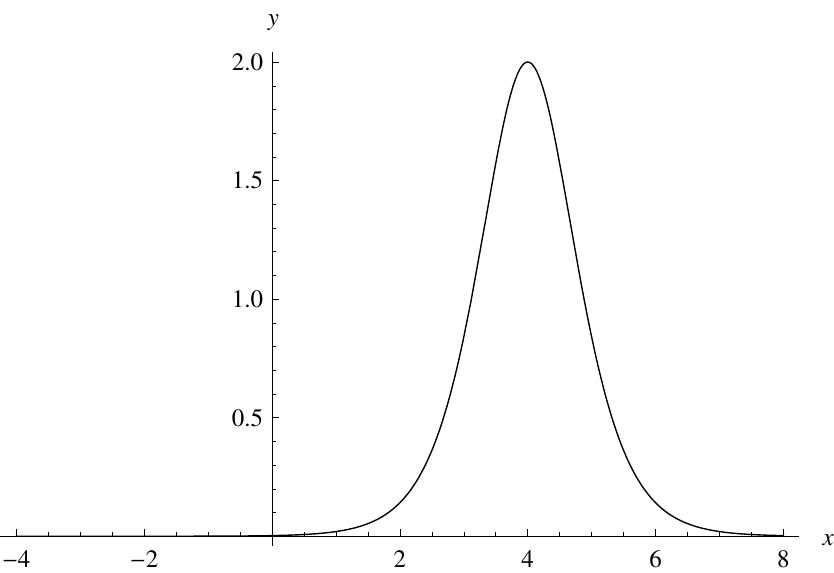}
 \caption{Soliton solution for $\kappa=1$, $t=1$, $x_0=0$, $y=-u(x,1)$}
\end{minipage}
\hfill
\begin{minipage}[b]{0.4\textwidth}
\includegraphics[width=\textwidth]{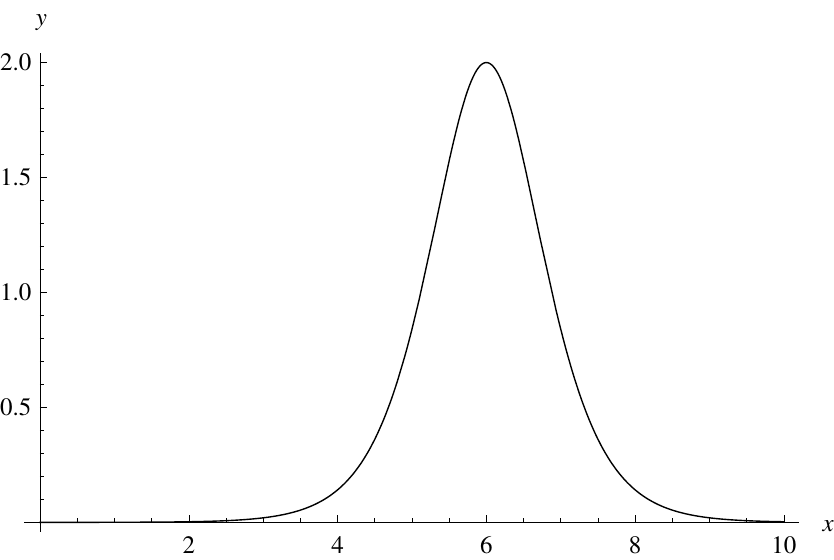}
 \caption{Soliton solution for $\kappa=1$, $t=1.5$, $x_0=0$, $y=-u(x,1.5)$}
\end{minipage}
\end{figure}
\par The algebraic method followed in deriving Eq. (9) can be generalized to include a reflectionless potential with $N$ bound states \cite{10}. In fact, a reflectionless potential with $N$ bound states corresponds, through the inverse scattering transform, to a pure $N$ soliton solution of the Korteweg-de Vries equation. To illustrate this we suppose that the bound states have parameters $\kappa_1,\; \kappa_2,...,\;\kappa_N$  and $c_1,\;c_2,...,\;c_N$. Then the kernel of the Gel'fand-Levitan equation is represented by 
\begin{equation}
 B(x+y,t)=\sum_{n=1}^Nc_n^2(t)e^{-\kappa_n(x+y)}=\sum_{n=1}^Nf_n(x,t)g_n(y).
\end{equation}
Here
\begin{equation}
 f_n(x,t)=c_n^2(t)e^{-\kappa_nx}\;\;\;\mbox{and}\;\;\;g_n(y)=e^{-\kappa_ny}.
\end{equation}
Then assuming a separable form of the kernel as 
\begin{equation}
 K(x,y,t)=\sum_{n=1}^Nk_n(x,t)g_n(y)
\end{equation}
we substitute (11) and (13) in the Gel'fand-Levitan  equation to write
\begin{equation}
 \sum_{n=1}^Nk_n(x,t)g_n(y)+\sum_{n=1}^Nf_n(x,t)g_n(y)+\int_x^\infty[\sum_{n=1}^Nk_n(x,t)g_n(s)\sum_{n=1}^Nf_n(s,t)g_n(y)]ds=0.
\end{equation}   
Making use of Eq. (12)  in Eq. (14) and equating the coefficient of $g_n(y)$ to zero we deduce  
\begin{equation}
 k_m+f_m+c_m^2\sum_{n=1}^Nk_m\frac{e^{-(k_m+k_n)x}}{(k_m+k_n)}.
\end{equation}
In writing Eq. (15) we have made use of the middle term of Eq. (11). In Matrix form Eq. (15) reads
\begin{equation}
 Mk+f=0,
\end{equation}
where $k$ and $f$  are column vectors with entries $k_n$ and $f_n$ respectively, and $M$ is an $N\times N$ square matrix with elements
\begin{equation}
 M_{ij}=\delta_{ij}+c_i^2(t)\frac{e^{-(\kappa_i+\kappa_j)x}}{\kappa_i+\kappa_j}.
\end{equation}
The matrix equation (16) can be solved so as to write
\begin{equation}
 K(x,y,t)=\frac{\partial}{\partial x}\ln \det M,
\end{equation}
which finally gives the $N$ soliton solution 
\begin{equation}
 u(x,t)=-2\frac{\partial^2}{\partial x^2}\ln\det M.
\end{equation}
We shall now illustrate the application of the above results by constructing (A) two-soliton solution, and (B) three-soliton solution of the KdV equation.
\subsection{The two-soliton solution of the KdV equation}
Construct the two-soliton solution of the KdV equation for the initial value given by  
\begin{equation}
 u(x,0)=-6\sec h^2x.
\end{equation}
For $i,\; j=1,\;2$, Eq. (17) can be judiciously used to construct the expression
\begin{equation}
 \det M=1+3e^{8t-2x}+3e^{64t-4x}+e^{72t-6x}.
\end{equation}
In writing (21) we have used $\kappa_1=1$, $\kappa_2=2$, $c_1=\sqrt{6}e^{4t}$ and $c_2=\sqrt{12}e^{32t}$. From Eq. (19) and Eq. (21) we have 
\begin{equation}
u(x,t)=-\frac{24e^{8t+2x}(e^{128t}+e^{8x}+4e^{72t+2x}+6e^{64t+4x}+4e^{56t+6x})}{(e^{72t}+e^{6x}+3e^{64t+2x}+3e^{8t+4x})^2}. 
\end{equation}
At $t=0$ the plot of Eq. (22) as a function of $x$ is shown in Fig. 5. The soliton in Fig. 5 is in exact agreement with that found from the formula in Eq. (20). The important point to note here is that the two-soliton solution at $t=0$ coalesce to give rise to a single soliton. We shall now show that as time goes on we find solitons separated by some distance that increases with time. In Fig. 6 we display the soliton solution from Eq. (22) as a function of $x$ for $t=0.5$.
\begin{figure}[h]
 \centering
\begin{minipage}[b]{0.4\textwidth}
\includegraphics[width=\textwidth]{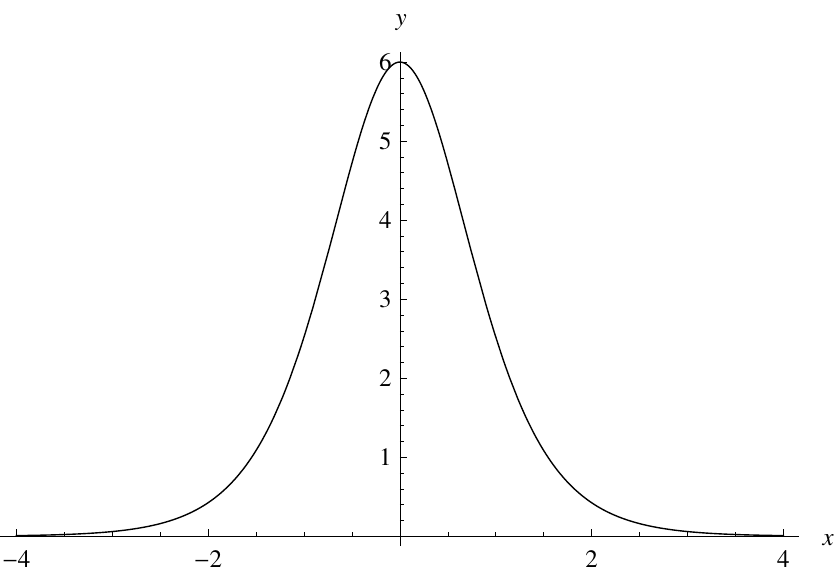}
 \caption{$u(x,0)$ as a function of $x$ from Eq. (22), $y=-u(x,0)$}
\end{minipage}
\hfill
\begin{minipage}[b]{0.4\textwidth}
\includegraphics[width=\textwidth]{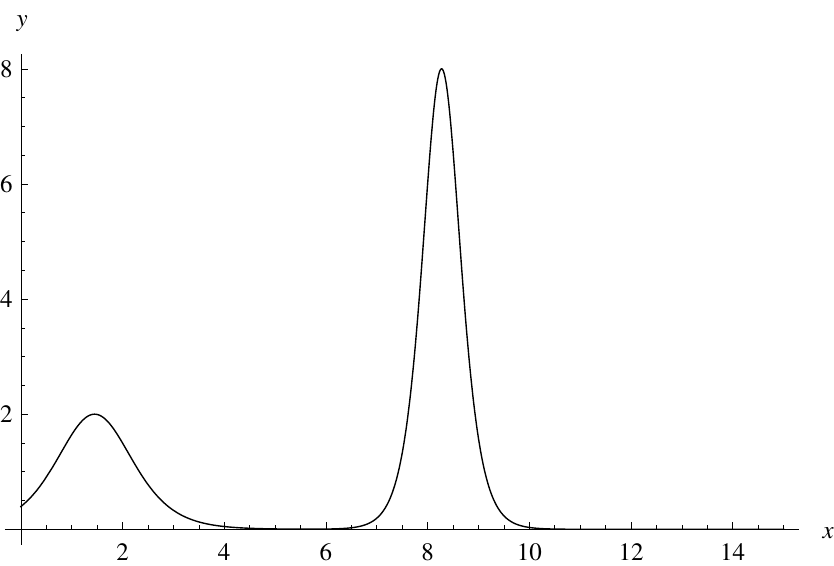}
 \caption{$u(x,0.5)$ as a function of $x$ from Eq. (22), $y=-u(x,0.5)$}
\end{minipage}
\end{figure}
\begin{figure}[h]
 \centering
\begin{minipage}[b]{0.4\textwidth}
\includegraphics[width=\textwidth]{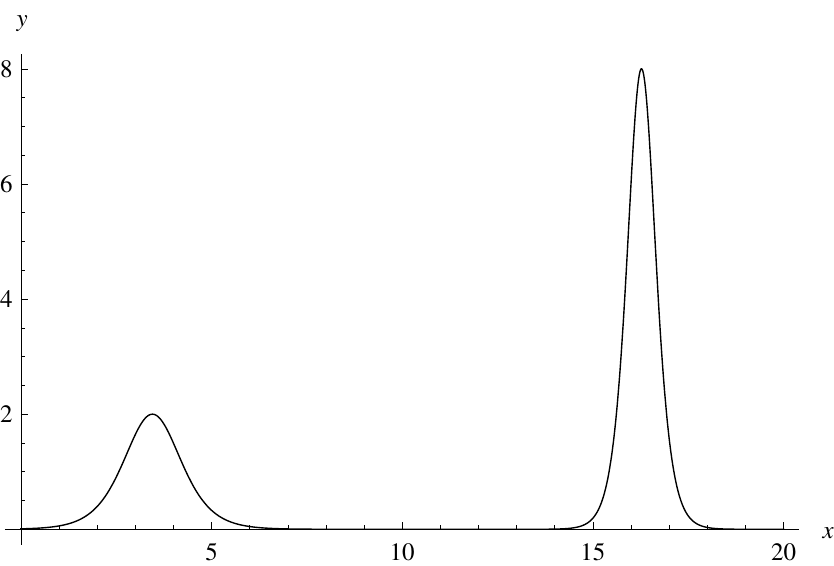}
 \caption{$u(x,1)$  as a function of $x$  from Eq. (22), $y=-u(x,1)$}
\end{minipage}
\hfill
\begin{minipage}[b]{0.4\textwidth}
\includegraphics[width=\textwidth]{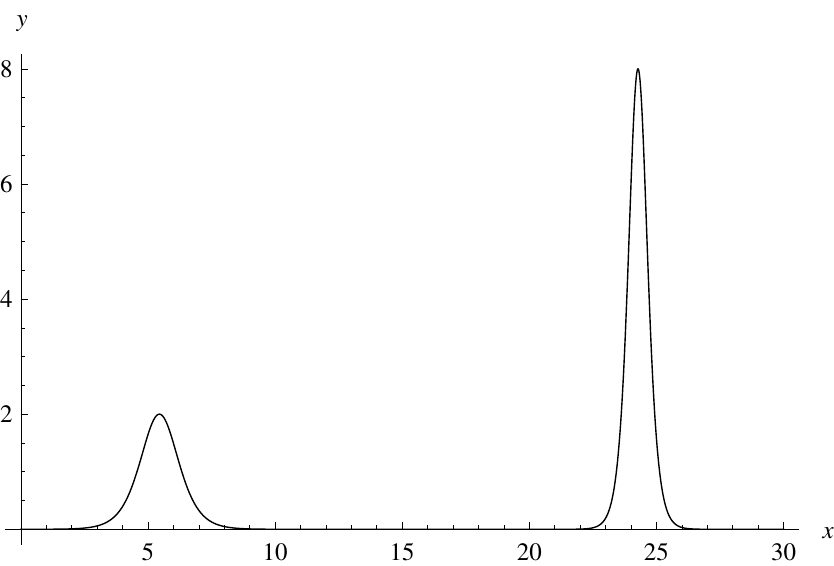}
 \caption{$u(x,1.5)$ as a function of $x$  from Eq. (22), $y=-u(x,1.5)$}
\end{minipage}
\end{figure}
The curve in this figure shows that the juxtaposed soliton solution of Fig. 5 transforms into a two separate solitons for $t>0$. It will be interesting to know how the solitons displayed in Fig. 6 behave during time evolution. In order that we plot in Figs. 7 and 8 and $u(x,1)$ and $u(x,1.5)$ as a function of $x$. From the curves in Figs. 6 - 8 it is clear that one soliton is taller than the other and during time evolution the tall soliton moves faster than the short one. As usual the coordinate for the centre of the single hump soliton in Fig. 5 is as predicted from Eq. (20). Coordinates of the peaks of the short and tall solitons in Figs. 6 - 8 are $\{(1.5,\;2),\;(8.3,8)\}$, $\{(3.4,\;2),\;(16.3,\;8)\}$ and $\{(5.5,\;2),\; (24.4,\;8)\}$. In each of these entries, the numbers in the round brackets encapsulated inside curly brackets denote the coordinates of the short and tall solitons. We shall follow this convention throughout. The area under the curve in Fig. 5 is 12 while in Fig. 6 the area of the curves representing short and tall soliton are 4 and 8 respectively. This implies that the energy of the juxtaposed solitons is redistributed into the component solitons such that the tall component carries two-third of the total energy while the short one carries only one-third of the total energy. The energy distribution remains constant during time evolution of the soliton.
\subsection{The three-soliton solution of the KdV equation}
Construct the three-soliton solution of the KdV equation for the initial value given by
\begin{equation}
u(x,0)=-12\sec h^2x.
\end{equation}
As with the problem in (A), for $i,\;j=1,\;2,\;3$, we can obtain from Eq. (17)
\begin{equation}
\det M=1+e^{288t-12x}+6e^{280t-10x}+15e^{224t-8x}+10e^{72t-6x}+10e^{216t-6x}+15e^{64t-4x}+6e^{8t-2x}
\end{equation}
such that
\begin{equation}
 u(x,t)=n/d
\end{equation}
with
\begin{equation}
\begin{aligned}
 n=&-48e^{8t+2x}(e^{560t}+e^{20x}+30e^{16(4t+x)}+15e^{16(13t+x)}+252e^{10(28t+x)}+50e^{8(36t+x)}+135e^{8(42t+x)}\\
& +25e^{8(54t+x)}+15e^{4(88t+x)}+10e^{504t+2x}+30e^{496t+4x}+80e^{344t+6x}+40e^{488t+6x}+25e^{128t+12x}\\
& +135e^{244t+12x}+50e^{272t+12x}+40e^{72t+14x}+80e^{216t+14x}+10e^{56t+18x})
\end{aligned}
\end{equation}
and
\begin{equation}
 d=(e^{288t}+e^{12x}+15e^{8(8t+x)}+10e^{6(36t+x)}+15e^{4(56t+x)}+6e^{280t+2x}+10e^{72t+6x}+6e^{8t+10x})^2.
\end{equation}
In Fig. 9 we display $–u(x,0)$ as a function of $x$ for the three-soliton solution as obtained from Eq. (25). As expected here we have three juxtaposed solutions so as to look like single-soliton. In fact the plot in Fig. 9 coincides with that obtained from Eq. (23). In Figs. 10 - 12 we portray the positions of the moving three-soliton solutions from Eq. (25). In each of these figures the position of the shorter soliton lies near the origin of the coordinates while ones with bigger heights are situated at $x>>0$. In particular, the tallest soliton occupies the furthest position.
\begin{figure}[h]
 \centering
\begin{minipage}[b]{0.4\textwidth}
\includegraphics[width=\textwidth]{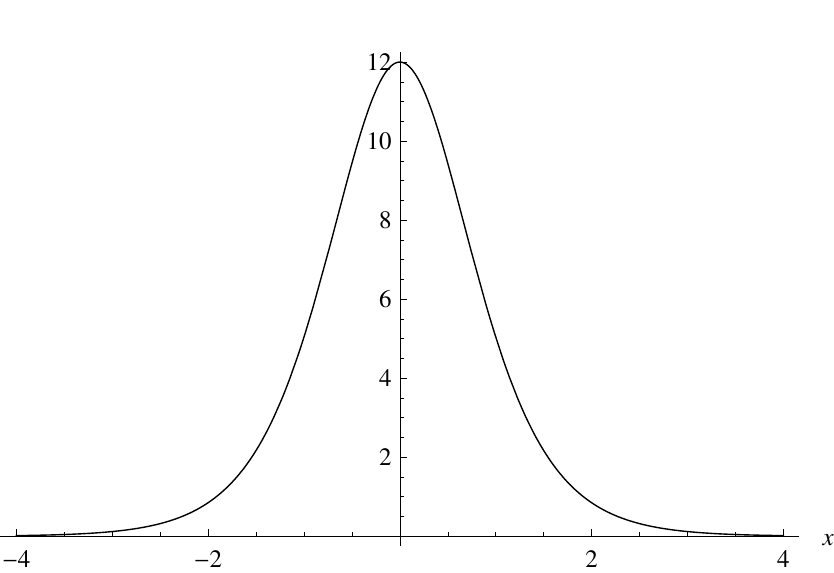}
 \caption{$u(x,0)$ as a function of $x$ from Eq. (25), $y=-u(x,0)$}
\end{minipage}
\hfill
\begin{minipage}[b]{0.4\textwidth}
\includegraphics[width=\textwidth]{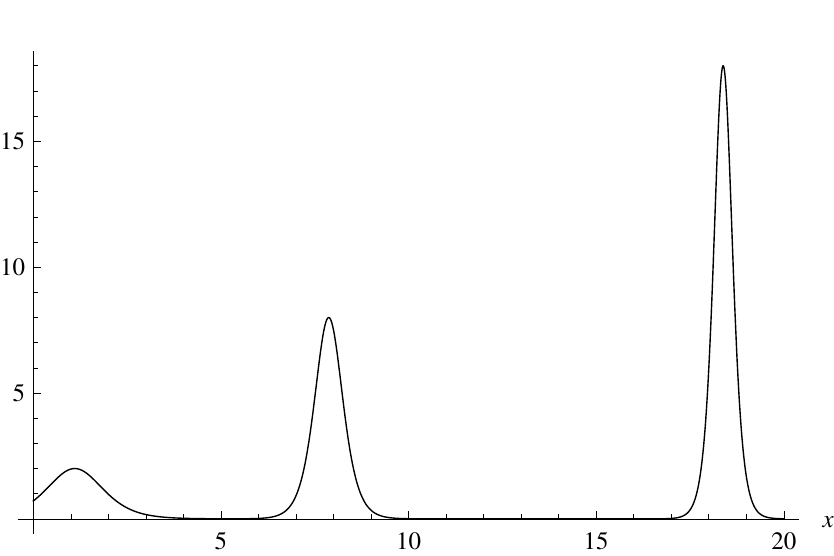}
 \caption{$u(x,0.5)$ as a function of $x$ from Eq. (25), $y=-u(x,0.5)$}
\end{minipage}
\end{figure}
\begin{figure}[h]
 \centering
\begin{minipage}[b]{0.4\textwidth}
\includegraphics[width=\textwidth]{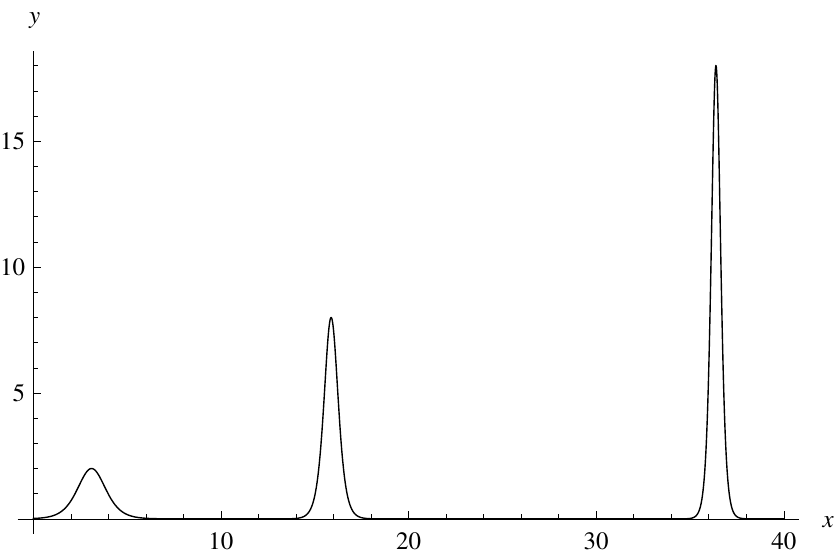}
 \caption{$u(x,1)$  as a function of $x$  from Eq. (25), $y=-u(x,1)$}
\end{minipage}
\hfill
\begin{minipage}[b]{0.4\textwidth}
\includegraphics[width=\textwidth]{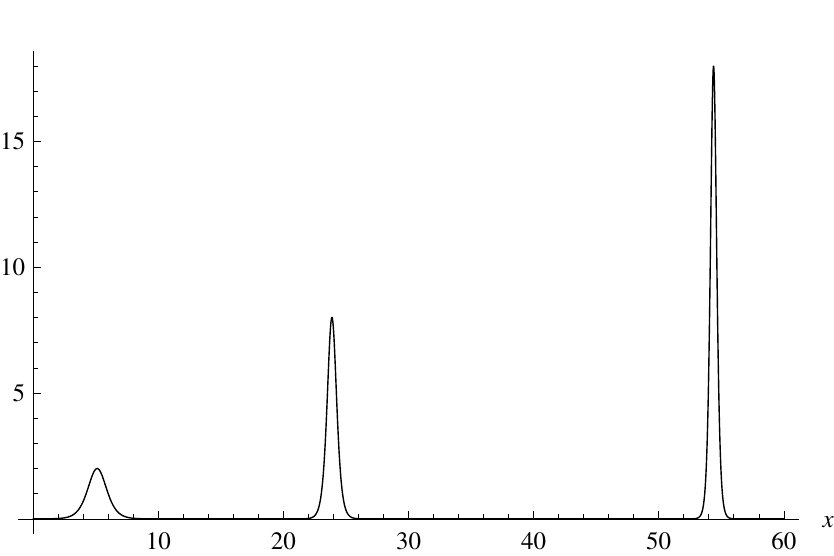}
 \caption{$u(x,1.5)$ as a function of $x$  from Eq. (25), $y=-u(x,1.5)$}
\end{minipage}
\end{figure}
The coordinates of the peaks of the solitons in the above three figures are given by $\{(1.2,2),\;(7.9,9),\;(18.8,17.9)\}$,  $\{(3.1,2),\;(16,9),\;(36.4,17.9)\}$, $\{(5,2),\;(24,9),\;(54.5,17.9)\}$. From the result presented for the centres of the three-soliton solutions one can easily verify that the taller soliton moves with greater velocity. Figures 10 - 12 also show that the taller solitons are narrower. As expected at $t=0$ the area under the curves in Fig. 9 is 24. For $t>0$ the energy of the juxtaposed soliton is redistributed into three solitons with areas 4, 8 and 12. Understandably, these areas remain constant during time evolution of the soliton.
\section*{3. Method of Lax for associating nonlinear differential equations with linear operators}
The success of inverse scattering method as applied to the KdV equation naturally raised the question of whether other nonlinear equations could be solved by analogous methods. This problem was considered by Lax \cite{5} who demonstrated in a seminal paper that a nonlinear evolution equation
\begin{equation}
 F(x,t,u,...)=0
\end{equation}
is integrable if there exists a pair of operators such that they yield the equation when they commute. Let us illustrate the method of Lax by particular attention to the KdV equation. In the inverse spectral method the KdV equation is solved by using the spectral problem
\begin{equation}
 L\psi=\lambda\psi.
\end{equation}
Appropriately the operator $L$ involves $u(x,t)$ and $\lambda_t=0$. In the Lax's method one introduces an auxiliary spectral problem defined by
\begin{equation}
 \psi_t=b\psi.
\end{equation}
Clearly, the operator $B$ characterizes the change in the eigen vector with respect to the parameter $t$ which in the KdV equation corresponds to time. We now differentiate with respect to $t$ and make use of Eq. (30) together with $\lambda_t=0$ and this obtain 
\begin{equation}
 L_t=[B,L],
\end{equation}
where the commutator $[B,L]=BL-LB$. For the KdV equation the operators $L$ and $B$, often called the Lax pair, can be identified by the use of the inverse scattering technique to write
\begin{subequations}
\begin{equation}
 L=-\frac{\partial^2}{\partial x^2}+u
\end{equation}
\mbox{and}
\begin{equation}
 B=-4\frac{\partial^3}{\partial x^3}+6u\frac{\partial}{\partial x}+3\frac{\partial u}{\partial x}.
\end{equation}
\end{subequations}
For the operator in Eq. (32a)
\begin{subequations}
\begin{equation}
L_t=u_t. 
\end{equation}
\mbox{Also introducing an arbitrary function $\phi=\phi(x,t)$ we can compute $BL\phi$ and $LB\phi$ to get}
\begin{equation}
[BL-LB]\phi=[-\frac{\partial^3u}{\partial x^3}+6u\frac{\partial u}{\partial x}]\phi
\end{equation}
\mbox{so that}
\begin{equation}
 [BL-LB]=-\frac{\partial^3u}{\partial x^3}+6u\frac{\partial u}{\partial x}
\end{equation}
\end{subequations}
From Eq. (31) and Eq. (33) we get the KdV equation.
\par The KdV equation represents a generic model for studying weakly nonlinear waves. Physically, this equation models surface waves with small amplitude and long wavelength on shallow water. The equation involves a balance between weak nonlinearity and linear dispersion (the second and third terms of Eq. (1). The Miura transformation \cite{11}
\begin{equation}
 u=\frac{\partial v}{\partial x}+v^2
\end{equation}
converts the KdV equation to another nonlinear equation given by
\begin{equation}
 \frac{\partial v}{\partial t}-6v^2\frac{\partial v}{\partial x}+\frac{\partial^3v}{\partial x^3}=0.
\end{equation}
The dispersive term of Eq. (35) is similar to that of the KdV equation but the middle term shows that this equation is more nonlinear than the KdV equation. Eq. (35) is known as the modified KdV equation and is often denoted as mKdV equation. The modified KdV equation manifests itself in diverse areas of physics. For example, it appears in the context of electromagnetic waves in size-quantized films, van Alfven waves in collisionless plasma, phonons in anharmonic lattice, ion acoustic solitons and many more \cite{12}.
\par From Eq. (32) and Eq. (34) the Lax pair for the mKdV equation can be written as
\begin{subequations}
\begin{equation}
 L=-\frac{\partial^2}{\partial x^2}+\frac{\partial v}{\partial x}+v^2
\end{equation}
\mbox{and}
\begin{equation}
 B=-4\frac{\partial^3}{\partial x^3}+6(\frac{\partial v}{\partial x}+v^2)\frac{\partial}{\partial x}+3\frac{\partial}{\partial x}(\frac{\partial v}{\partial x}+v^2).
\end{equation}
\end{subequations}
As in the case of KdV equation Eq. (36a) and Eq. (36b) can be inserted in the Lax equation (31) to get
\begin{equation}
 (\frac{\partial}{\partial x}+2v)(\frac{\partial v}{\partial t}-6v^2\frac{\partial v}{\partial x}+\frac{\partial^3v}{\partial x^3})=0.
\end{equation}
Clearly, the second bracketed term of Eq. (37) must be equal to zero. Thus we identify that $v$  must be a solution of the mKdV equation. 
\par In the above we have obtained the Lax pair of the mKdV equation from that of the KdV equation by using a simple mapping between $u$  and $v$ as given in Eq. (34). But the general problem of finding Lax pairs for nonlinear evolution equations is quite complicated. Without going into such details we present below the result of Lax pairs for a number of physically important equations. As opposed to the third-order equations as considered above we shall consider now first a fifth-order equation 
\begin{equation}
 \frac{\partial u}{\partial t}+5u^2\frac{\partial u}{\partial x}+5u\frac{\partial^3 u}{\partial x^3}+5\frac{\partial u}{\partial x}\frac{\partial^2 u}{\partial x^2}+\frac{\partial^5 u}{\partial x^5}=0
\end{equation}
given by Sawada and Kotera (SK) \cite{13}. The Lax pair for Eq. (38) is given by 
\begin{subequations}
\begin{equation}
 L=\frac{\partial^3}{\partial x^3}+u\frac{\partial}{\partial x}
\end{equation}
\mbox{and}
\begin{equation}
 B=9\frac{\partial^5}{\partial x^5}+15u\frac{\partial^3}{\partial x^3}+15\frac{\partial u}{\partial x}\frac{\partial}{\partial x}+(5u^2+10\frac{\partial^2 u}{\partial x^2}).
\end{equation}
\end{subequations}
Although the algebra is little lengthy, using Eq. (39a) and Eq. (39b) one can get Eq. (38). In the following we provide the Mathematica program to compute the KdV, mKdv and SK equations from their Lax pairs.
\vskip0.25cm
\par {\bf KdV equation}\\
f:= $\psi$[x,t]\\g:=u[x,t]\\LaxL[f$_{-}$]:=-D[f,\{x,2\}]+g\;f\\LaxB[f$_{-}$]:=-4\;D[f,\{x,3\}]+6\;g\;D[f,x]+3\;f\;D[g,x]\\
Simplify[LaxB[LaxL[f]]- LaxL[LaxB[f]]-D[g,t]\;f]
\vskip0.25cm
\par {\bf mKdV equation}\\
f:= $\psi$[x,t]\\g:=v[x,t]\\LaxL[f$_{-}$]:=-D[f,\{x,2\}]+(D[g,x]+g\textasciicircum 2)\;f\\LaxB[f$_{-}$]:=-4\;D[f,\{x,3\}]+6 (D[g,x]+g\textasciicircum2)\; D[f,x]+3\;D[(D[g, x]+g\textasciicircum2),x] f\\Simplify[LaxB[LaxL[f]]-LaxL[LaxB[f]]-D[(D[g,x]+ g\textasciicircum2),t]\;f]
\vskip0.25cm
\par {\bf Sawada and Kotera equation}\\
f:= $\psi$[x,t]\\g:=v[x,t]\\LaxL[f$_{-}$]:=D[f,\{x,3\}+g D[f,x]]\\LaxB[f$_{-}$]:=9 D[f,\{x,5\}]+15 g D[f,\{x,3\}]+15 D[g,x] D[f,\{x,2\}]+(5 g\textasciicircum2+10 D[g,\{x,2\}]) D[f,x]\\Simplify[LaxB[LaxL[f]]-LaxL[LaxB[f]]-D[g,t] D[f,x]]
\vskip0.5cm
\par In the following we shall present results for the Lax pairs of a number of physically important nonlinear evolution equations.
\par The NLS equation: The nonlinear Schr\"{o}dinger (NLS) equation is given by \cite{12}
\begin{equation}
 iu_t+u_{xx}+2\lambda\lvert u\rvert^2u=0
\end{equation}
with
\begin{equation}
 u(x,0)=u_0(x)\in S(\Re).
\end{equation}
Here $S(\Re)$  denotes Schwartz class of rapidly decaying functions. This classical equation (40), for $\lambda=0$ gives the well known Schr\"{o}dinger equation in quantum mechanics. When $\lambda=1$ we obtain the focusing NLS equation and for $\lambda=-1$, the defocusing NLS equation.The equation have applications in various fields such as acoustics, optics as well Bose-Einstein condensates \cite{14}. In addition to the KdV equation, the equation in Eq. (40) also represents the canonical examples of the (1+1)-dimensional partial differential equations. For each of the focusing and defocusing NLS equations there exists an inverse scattering transform. The focusing NLS equation associated with the system of linear-ordinary differential equations 
\begin{subequations}
 \begin{equation}
  \frac{d\xi}{dx}=-i\kappa\xi+u\eta
 \end{equation}
\mbox{and}
\begin{equation}
 \frac{d\eta}{dx}=i\kappa\eta-u^*\xi
\end{equation}
\end{subequations}
is known as the Zakharov-Shabat system \cite{15}. Here asterisk denotes complex conjugation. The corresponding Lax pair is given by 
\begin{subequations}
\begin{equation}
 L=\left(
    \begin{array}{cc}
     i\partial_x & iu\\-iu^* & -i\partial_x
    \end{array}
 \right)
\end{equation}
\mbox{and}
\begin{equation}
 B=\left(
\begin{array}{cc}
   2i\partial_x^2+i |u|^2 & -2iu\partial_x-iu_x\\-2iu^*\partial_x-iu_x^* & -2i\partial_x^2-i|u|^2
\end{array}
\right).
\end{equation}
\end{subequations}
For the defocusing nonlinear Schr\"{o}dinger equation
\begin{equation}
 u_t+u_{xx}-2|u|^2u=0
\end{equation}
the associated linear differential equations are
\begin{subequations}
\begin{equation}
 \frac{d\xi}{dx}=-i\lambda\xi+u(x,t)\eta
\end{equation}
\mbox{and}
\begin{equation}
 \frac{d\eta}{dx}=i\lambda\eta-u(x,t)^*\xi
\end{equation}
\end{subequations}
with the corresponding Lax pair
\begin{subequations}
\begin{equation}
 L=\left(
    \begin{array}{cc}
     i\partial_x & -iu\\iu^* & -i\partial_x
    \end{array}
 \right)
\end{equation}
\mbox{and}
\begin{equation}
 B=\left(
\begin{array}{cc}
   2i\partial_x^3-i|u|^2 & -2i\partial_x-iu_x\\2iu^*\partial_x+iu^* & -2i\partial_x^2+iu^*
\end{array}
\right).
\end{equation}
\end{subequations}
The sine-Gordon equation: This hyperbolic partial differential equation given by \cite{16}
\begin{equation}
 u_{xt}=\sin u
\end{equation}
is associated with the linear system
\begin{subequations}
\begin{equation}
 \frac{d\xi}{dx}=-i\lambda\xi-\frac{1}{2} u_x(x,t)\eta
\end{equation}
\mbox{and}
\begin{equation}
 \frac{d\eta}{dx}=i\lambda\eta+\frac{1}{2}u_x(x,t)^*\xi.
\end{equation}
\end{subequations}
The corresponding Lax pair is given by
\begin{subequations}
\begin{equation}
 L=\left(
    \begin{array}{cc}
     i\partial_x & \frac{iu_x}{2}\\\frac{iu_x}{2} & -i\partial_x
    \end{array}
 \right)
\end{equation}
\mbox{and}
\begin{equation}
 B=\frac{1}{8}\left(\int_{-\infty}^x-\int_x^\infty\right)dy\left[\cos\left(\frac{u(x,t)+u(y,t)}{2}\right)\left(
\begin{array}{cc}
 1 & 0\\0 & 1
\end{array}
\right)
+\sin\left(\frac{u(x,t)+u(y,t)}{2}\right)\left(
\begin{array}{cc}
 0 & -1\\1 & 0
\end{array}
\right)\right]
\end{equation}
\end{subequations}
\section*{4. AKNS hierarchy and zero-curvature representation}
The so-called Lax pair representaton for the KdV equation was given by Lax himself \cite{5}. However, a similar Lax pair for the standard nonlinear Schr\"{o}dinger (both focusiong and defocusing) was found by Zakharov and Shabat \cite{15} and its integrability was established  by Zakharov and Manakov \cite{17}. Simultaneously, with the works in refs. 11 and 13, Ablowitz, Kaup, Newell and Segur \cite{6} introduced a more general kind of Lax pairs and thus provided a systematic method to find new integrable systems which constitute the socalled AKNS hierarchy. In the approach of Ablowitz et. al., one begins by introducing a $2\times 2$ linear eigen-value problem 
\begin{subequations}
 \begin{equation}
  \frac{\partial\Phi}{\partial x}=U\Phi
 \end{equation}
\mbox{and}
\begin{equation}
 \frac{\partial\Phi}{\partial t}=V\Phi
\end{equation}
\end{subequations}
for the nonlinear evolution equation. Here $U$ and $V$ are traceless matrices given by
\begin{subequations}
\begin{equation}
 U=\left(
    \begin{array}{cc}
     -\eta(x,t) & q(x,t)\\r(x,t) & \eta(t)
    \end{array}
 \right)
\end{equation}
\mbox{and}
\begin{equation}
 V=\left(
    \begin{array}{cc}
     A(x,t) & B(x,t)\\C(x,t) & -A(x,t)
    \end{array}
 \right).
\end{equation}
\end{subequations}
The element $\eta(t)$ is the time dependent spectral parameter and $q(x,t)$, $r(x,t)$, $A(x,t)$, $B(x,t)$ and $C(x,t)$ are real functions of space and time variables. From the integrability condition of Eq. (50a) and Eq. (50b) i.e. $\Phi_{xt}=\Phi_{tx}$ it is rather straightforward to see that the traceless matrices $U$ and $V$ are constrained by the zero curvature condition \cite{19}
\begin{equation}
 U_t-V_t+[U,V]=0.
\end{equation}
From Eqs. (31) and (52), it is apparent that $U$ and $V$ represent the matrix Lax pair of the nonlinear equation. The differential representation of the Lax pairs as given in Eq. (32), Eq. (36) or Eq. (39) are not symmetric either in $\partial_x$ or in $u$. As opposed to this the matrix Lax pair appears to be symmetric. The price we pay to bring in this symmetry, whatsoever, is that $U$ and $V$ are now square matrices, and $\Phi$, a column vector. It may be of some interest to know why Eq. (52) has been given the name zero curvature condition and or representation.
\par In general theory of relativity the gravitational field is associated with space-time curvature \cite{19}. It turns out that the electromagnetic field is also associated with curvature not of space-time as those in the relativity theory but of an internal space defined by $U(1)$ principal bundle over space-time \cite{20}. It is well known that the electric and magnetic fields can be packaged in the so-called field strength tensor such that $F_{0i}=E_i/c$  and $F_{ij}=\sum_k\epsilon_{ijk}B_k$ for $1\leq i,\;j,\;k\leq3$, where $c$ is the speed of light. Thus the field strength is a measure of curvature. Specializing to $(1+1)$-dimension written as $x^0=t,\;x^1=x$ and introducing the scalar and vector potentials $A_0$ and $A_1$ we have $F_{01}=\partial_tA_1-\partial_xA_0$. In non-abelian gauge theories relevant to the strong and weak interactions, $A_0$ and $A_1$ become square matrices and the field strength acquires an extra commutator term such that $F_{01}=\partial_tA_1-\partial_x A_0+[A_1,A_0]$. Now if we use the mapping $A_1\rightarrow U$  and $A_0\rightarrow V$ we see that the consistency condition (52) is similar to that of $F_{01}$, the zero curvature of the non-abelian gauge field. Hence Eq. (52) has been given the name zero curvature condition.
\par For the KdV equation the matrix Lax pair of AKNS is given by 
\begin{subequations}
\begin{equation}
U=\left(
    \begin{array}{cc}
     0 & u-\lambda\\1 & 0
    \end{array}
 \right)
\end{equation}
\mbox{and}
\begin{equation}
V=\left(
    \begin{array}{cc}
     u_x & -4\lambda^2+2\lambda u+2u^2-u_{xx}\\4\lambda+2u & -u_x
    \end{array}
 \right).
\end{equation}
\end{subequations}
The matrix Lax pair representation for the Zakharov-Shabat system or the focusing NLS equation can be written as
\begin{subequations}
\begin{equation}
U=\left(
    \begin{array}{cc}
     -i\lambda & u\\-u^* & i\lambda
    \end{array}
 \right)
\end{equation}
\mbox{and}
\begin{equation}
V=\left(
    \begin{array}{cc}
     -2i\lambda^2+i|u|^2 & 2\lambda u+iu_x\\-2\lambda u^*+iu_x^* & 2i\lambda^2-i|u|^2
    \end{array}
 \right).
\end{equation}
\end{subequations}
The defocusing NLS equation is characterized by the matrix Lax pair
\begin{subequations}
\begin{equation}
U=\left(
    \begin{array}{cc}
     -i\lambda & u\\u^* & i\lambda
    \end{array}
 \right)
\end{equation}
\mbox{and}
\begin{equation}
V=\left(
    \begin{array}{cc}
     -2i\lambda^2-i|u|^2 & 2\lambda u+iu_x\\-2\lambda u^*+iu_x & 2i\lambda^2+i|u|^2
    \end{array}
 \right).
\end{equation}
\end{subequations}
The focusing mKdV equation corresponds to the matrix Lax pair
\begin{subequations}
\begin{equation}
U=\left(
    \begin{array}{cc}
     -i\lambda & u\\-u & i\lambda
    \end{array}
 \right)
\end{equation}
\mbox{and}
\begin{equation}
V=\left(
    \begin{array}{cc}
     -4i\lambda^3+2i\lambda u^2 & 4\lambda^2u+2i\lambda u_x-u_{xx}-2u^3 \\-4\lambda^2u+2i\lambda u_x+u_{xx}+2u^3 & 4i\lambda^3-2i\lambda u^2
    \end{array}
 \right).
\end{equation}
\end{subequations}
The matrix Lax pair for the sine-Gordon equation is given by
\begin{subequations}
\begin{equation}
U=\left(
    \begin{array}{cc}
     -i\lambda & -\frac{1}{2}u_x \\ \frac{1}{2}u_x & i\lambda
    \end{array}
 \right)
\end{equation}
\mbox{and}
\begin{equation}
V=\frac{i}{4\lambda}\left(
    \begin{array}{cc}
     \cos u & \sin u\\\sin u & -\cos u
    \end{array}
 \right).
\end{equation}
\end{subequations}
\section*{5. Geometry of the zero curvature condition}
In the above we have cited a few examples to point out that soliton equations satisfy the curvature condition.The relevant mathematical framework underlying this formalism is embedded in noncommutative geometry \cite{21}. In the noncommutative geometry an associative but not necessarily commutative algebra replaces the algebra of smooth functions on a manifold. A differential calculus on the algebra is then regarded as the most basic geometric structure on which further geometric concepts like connections can be defined. In the case of KdV, sine-Gordon and sinh-Gordon equations, one can find $SL(2,\Re)$ -connection 1-form (gauge potentials) such that the condition for vanishing curvature (or 'field strength')
\begin{equation}
 F=DA+AA=0
\end{equation}
is equivalent to the respective soliton equation. Thus there is a special geometric feature related to the integrability of nonlinear equations. We shall try to bring out some geometrical properties of the zero curvature representation of nonlinear equations relating to propagation of solitons in one-dimensional space. 
\par We begin by noting that for nonlinear differential equations, the zero curvature condition is obtained from the compatibility condition of Eq. (50a) and Eq. (50b). Equation (50a) implies the motion of the auxiliary function $\Phi(x,t)$ in the $(x,\;t)$ - plane \cite{22} in the $x$ direction with a square matrix. Similarly, Eq. (50b) refers to the motion of $\Phi(x,t)$ in the $(x,\;t)$ - plane in the $t$  direction with the square matrix $V$. Written explicitly, the integrability and of the compatibility condition reads
\begin{equation}
 \frac{\partial}{\partial x}\left(\frac{\partial}{\partial t}\Phi(x,t)\right)=\frac{\partial}{\partial t}\left(\frac{\partial}{\partial x}\Phi(x,t)\right).
\end{equation}
Equation (59) has a simple geometrical interpretation as this equation describes connections on a two-dimensional vector bundle over the $(x,\;t)$ - plane. More specifically, Eq. (50a) shows the 'parallel translation' of $\Phi(x,t)$ in the $x$ -direction while Eq. (50b) implies similar 'parallel translation' along the $t$ - direction. The matrices $U$ and $V$ stand for the so-called connection coefficients. A connection is defined to represent zero curvature if the parallel translation between two points is independent of the path connecting them. Thus the compatibility condition (59) represents the integrability of the nonlinear equation leading to soliton solutions.
\section*{6. Concluding remarks}
In this work we have followed a historical viewpoint to introduce the inverse spectral method as a very useful tool for solving a class of nonlinear evolution equations and emphasized that the equations solvable by this method support soliton solutions. All equations considered here can be represented by two linear differential operators often called the Lax pair. In addition to the differential form of the Lax pair, we have considered the matrix Lax pair which are constrained by zero curvature condition. Wherever possible, simple geometrical interpretation is sought for the results presented. The nonlinear differential equations soluble by the use of inverse scattering theory and endowed with Lax pair leading to zero curvature representation are often referred to as integrable systems. However, there in no strict definition of integrability.
\par  The most successful technique to investigate the integrability of nonlinear ordinary as well as partial differential equations consists in applying the so-called Painleve analysis for their singularity structure \cite{23}. This viewpoint asserts that a nonlinear partial differential equation is completely integrable if after similarity reduction it coincides with any of the six Painleve equations. Any nonlinear differential equation that passes the Painleve test possesses infinite number of conserved densities. An important physico-mathematical implication of having infinite number of conserved densities is that for a given integrable equation one can obtain a hierarchy of higher-order equations by taking recourse to the use of a hereditary operator \cite{24}.The higher-order equations so generated share the same spectral problem as was used to solve the mother equation in the hierarchy. The auxiliary spectral problem, however, changes as one goes along the hierarchy.
\par Zakharov abd Faddeev \cite{25} developed the Hamiltonian approach to integrability of nonlinear evolution equations in $(1+1)$- dimensional systems as considered in this paper. Almost simultaneously, Gardner \cite{26} interpreted the KdV equation as a completely integrable Hamiltonian system with $\partial_x$ as the relevant Hamiltonian operator. A significant development of the Hamiltonian theory is due to Magri \cite{27} who established that integrable Hamiltonian systems have an additional structure. They are bi-Hamiltonian i.e. Hamiltonian with respect to two different compatible Hamiltonian operators. One of the important criteria for a nonlinear partial differential equation to be integrable is that it should possess bi-Hamiltonian structure.

\end{document}